\documentclass{aa}  

\usepackage{graphicx}
\usepackage{txfonts}
\newcommand{\msec}[2]{$#1\mbox{$''\mskip-7.6mu.\,$}#2$}

\begin{document}

   \title{Rapid Angular Expansion of the Ionized Core of CRL~618}

   \subtitle{}

   \author{D. Tafoya\inst{1}
   	\and L. Loinard\inst{2} 
	\and J. P. Fonfr\'\i a\inst{3} 
	\and W. H. T. Vlemmings\inst{1} 
	\and I. Mart\'\i -Vidal\inst{1}
	\and G. Pech\inst{2}
          }

   \institute{Chalmers University of Technology, Onsala Space Observatory SE-439 92 Onsala, Sweden\\
         \and
             Centro de Radioastronom\'\i a y Astrof\'\i sica, UNAM, Apdo. Postal 3-72 (Xangari), 58089 Morelia, Michoac\'an, M\'exico\\
          \and
              Departamento de Estrellas y Medio Interestelar, Instituto de Astronom\'\i a, Universidad Nacional Aut\'onoma de M\'exico, 
Ciudad Universitaria, 04510, M\'exico City, M\'exico
             }

   \date{Received 2013; accepted 2013}

 
  \abstract
   {During the transition from the AGB to the planetary nebulae phase the circumstellar envelopes of 
   most of low and intermediate mass stars experience a dramatic change in morphology. CRL~618 is 
   an object that exhibits characteristics of both AGB and post-AGB star. It also displays a spectacular
   array of bipolar lobes with a dense equatorial region, which makes it an excellent object to study the 
   development of asymmetries in evolved stars. In the recent decades, an elliptical compact HII region 
   located in the center of the nebula has been seen to be increasing in size and flux. This seems to be 
   due to the ionization of the circumstellar envelope by the central star, and it would be indicating the 
   beginning of the planetary nebula phase for CRL~618. }
   {To determine the physical conditions under which the onset of the ionization the circumstellar 
   envelope of CRL~618 took place and the subsequent propagation of the ionization front. }
   {We analyzed interferometric radio continuum data at $\sim 5$ and 22~GHz from observations carried out at 
   seven epochs with the VLA. We traced the increase of the flux of the ionized region over a period of 
   $\sim 26$~years. We measured the dimensions of the HII region directly from the brightness distribution images to 
   determine the increase of its size over time. For one of the epochs we analyzed observations at six frequencies from which 
   we estimated the electron density distribution. We carried out model calculations of the spectral energy distribution at 
   two different epochs to corroborate our observational results.}
   {We found that the radio continuum flux and the size of the ionized region have been increasing monotonically in the last three decades. 
   The size of the major axis of the HII region shows a dependance with frequency, which has been interpreted as a result of a gradient 
   of the electron density in this direction. The growth of the HII region is due to the expansion of an ionized wind whose mass-loss rate 
   increased continuously for a period of $\sim 100$~years until a few decades ago, when the mass-loss rate experienced a sudden decline. 
   Our results indicate that the beginning of the ionization of the circumstellar envelope began 
   around 1971, which marks the start of the planetary nebula phase of CRL~618. }
   {}


\maketitle
%

\section{Introduction}

The transition from the Asymptotic Giant Branch (AGB) to the Planetary Nebula (PN) phase has long been known to be one of the 
briefest in stellar evolution. As a consequence, the number of Galactic objects in this transitionary stage (the so-called post-AGB stars 
and/or pre-planetary nebulae --PPNe) is small, and our understanding of the physical processes involved in the transition remains 
limited. It has been observed that during this transition the circumstellar envelopes (CSEs) of the AGB stars experience their 
most dramatic change in morphology. Therefore, it is important to characterize the physical conditions, as well as the main physical 
mechanisms, that predominate during the PPN phase to understand the complex shapes and structures that are seen in PNe. 

A unique opportunity to study this rapid transition is provided by the object CRL~618 (Westbrook Nebula, GL~618, (R)AFGL~618, 
IRAS~04395+3601). This source was discovered in the infrared sky survey of Walker \& Price (1975). Westbrook et al.\ (1975) 
studied this object thoroughly and suggested for the first time that it may be a nascent planetary nebula. Currently, in most of the 
literature it is classified as a carbon-rich bipolar PPN that is rapidly evolving toward the PN stage. 

The first optical images of this source showed a pair of small nebulosities lying along the east-west direction. The infrared emission 
was found to be located in the middle of the two visual nebulosities (Westbrook et al.\ 1975). Gottlieb \& Liller (1976) discovered that the 
$B$-magnitude of the nebulosities of CRL~618 increased from 18.8 to 16.5 in the period from 1940 to 1975. The first detection of 
radio continuum toward this source was reported by Wynn-Williams (1977) who detected emission at 5 and 15 GHz. Subsequently, 
Kwok \& Feldman (1981) discovered that CRL~618 was also exhibiting a brightening at radio wavelengths. These authors showed 
that in the period from 1977 to 1980 the emission at cm wavelengths increased by a factor of 2. They attributed the rise of the radio 
flux to the expansion of the ionized region, as a consequence of the increase of the temperature of the central star. 


High angular resolution images obtained with the Hubble Space Telescope revealed that the morphology of the optical emission of 
CRL~618 is dominated by a complex of collimated bipolar lobes extending along the E-W direction (Trammell 2000). These structures 
are composed of shock-excited gas with outflow velocities up to 200~km~s$^{-1}$, inclined by $30-45^\circ$ from the plane of the sky 
(Calvet \& Cohen 1978; Cernicharo et al. 1989; Neri et al. 1992; Cernicharo et al. 2001; S\'anchez-Contreras et al. 2002, 2004a,b; Pardo et al. 2005). 
In the center of the complex of lobes lies a compact HII region whose ionization is maintained by a 
B0V star with a temperature of $\simeq 27000-32000$~K, a mass of $\simeq 0.8$~M$_\odot$, and a luminosity of $\simeq 1-4\times 10^4$~L$_\odot$ 
(Westbrook et al. 1975; Calvet \& Cohen 1978; Kaler et al. 1978; Schmidt \& Cohen 1981; Kwok \& Bignell 1984; Goodrich et al. 1991; Knapp et al. 1993). 
The spectrum of this region is 
different from that of the lobes, indicating that the gas is completely photoionized with an electronic density  n$_{\rm e} \sim10^{5}$~-~10$^{7}$~cm$^{-3}$. 
The expansion velocity of the gas in this region seems to be v$_{\rm exp}\sim 30$~km~s$^{-1}$, much slower than that of the shocked lobes 
(S\'anchez-Contreras et al. 2002). 

In this article, we will concentrate on the central HII region of CRL~618. Interferometric radio images of this region revealed that the radio emission has a 
more or less elliptical brightness distribution whose size was found to be about \msec{0}{12}~$\times$~\msec{0}{36} at $\sim 22$~GHz (Kwok \& Bignell 1984). 
While the major axis (East-West direction) of the ionized region increases as a function of the observation wavelength, the minor axis (N-S direction) remains 
constant. This indicates that the HII region is ionization-bound in the North-South direction, suggesting an enhancement of the neutral material 
in this direction, which is consistent with the presence of a dense molecular torus (S\'anchez-Contreras \& Sahai 2004; S\'anchez-Contreras et al. 2004). 
As mentioned above, the radio continuum exhibited a considerable increase of flux since it was detected for the first time until around the end of 1980. 
Subsequent observations reported that the emission at cm wavelengths showed no significant increase (Mart\'in-Pintado et al. 1988; 1993; 1995). 

The millimeter emission of the compact HII region has also been found to be changing its flux with time (Mart\'{\i}n-Pintado et al.\ 1988; Walmsley 
et al.\ 1991; Knapp et al.\ 1993). However, the exact nature and significance of that rise of the flux remains unclear. Mart\'in-Pintado et al. (1988) suggested that the 
growth of the mm emission that they observed could be explained in terms of an ionized wind whose mass-loss rate has been increasing in the last decades. 
However, a recent decrease in the mm continuum flux suggests that the mass-loss rate has slowed down in the past few years. These authors 
modeled the spectral energy distribution and the recombination lines,  assuming a $r^{-2}$ power-law for the electron density. From the radio 
recombination line emission they derived an electron temperature for the ionized gas of $T_{\rm e}\sim 13000$~K. The value for the electron 
density in the center of the HII region that fitted their data best was on the order of 10$^{7}$~cm$^{-3}$, and the expansion velocity of the ionized 
gas was found to be $\sim 20$~km~s$^{-1}$. 

In this work, we present the results of analyzing radio continuum observations spanning more than 25 years. We show
that the radio continuum emission has continued its rise since it was discovered, and reveal for the first time 
the evolution of the expansion of the HII region by directly measuring the change of its size over time. The details of the
observations and data calibration and reduction are presented in \S2 of this paper. The measurements obtained from the data are reported in \S3. In \S4 we 
derive the physical parameters in the compact HII region and interpret our results. In \S5 we discuss the implications of our results for the nature 
of CRL~618 and the possible scenarios that could explain our observations. Finally in \S6 we present the conclusions of this work.

The distance to CRL~618 remains somewhat uncertain, with quoted values varying between 0.9 and 1.8 kpc (Goodrich et al.\ 1991; Schmidt \& Cohen 
1981; Knapp et al.\ 1993; S\'anchez-Contreras et al.\ 2004b). The value that has been used most recently in the literature is that obtained by Goodrich et al. (1991), 
$D=0.9$~kpc.  We will assume this value for the distance of CRL~618 throughout this article, unless stated otherwise. 


\section{Observations and Results}

In order to investigate the evolution of the structure of the HII region in CRL~618, we have analyzed seven Very Large Array 
(VLA) observations collected in the years 1982, 1983, 1990, 1992, 1995, 1998 and 2007. All of these data sets include observations
at the same wavelength (1.3 cm) and in the same configuration (A) of the array. Therefore they can be directly compared 
to one another to derive the evolution of the size and flux of the source at this specific wavelength ($\nu\approx 22$~GHz). In addition, 
the data obtained in 1983 includes observations at 5~GHz and those obtained in 1998 included observations at six frequency bands. 
The data calibration and the imaging were made using the AIPS package of the NRAO\footnote{The National Radio Astronomy Observatory 
is a facility of the National Science Foundation operated under cooperative agreement by Associated Universities, Inc.}. After the initial calibration,
using extragalactic compact sources as calibrators, we performed phase self-calibration to all the data sets. Subsequently clean 
images of the source were obtained using natural weighting (robust parameter = 5 in the task IMAGR of AIPS). The main parameters 
of the calibration and the resulting images are given in Tables 2 and 3. The typical size of the synthesized beam at 22~GHz was 
$\sim$\msec{0}{1}, thus we adopted this size as a common value for the circular clean beam used during the reconstruction of the 
images at this frequency of all the epochs. The flux density was measured by integrating the emission within a box that contained the 
whole source. Since the morphology of this source is rather simple, we measured its size by fitting a two-dimensional gaussian model to 
the emission (we verified that our main results and conclusions are not affected by adopting different fitting models to our data). The 
results from the fits are shown in Tables 4 and 6. For each observation we followed a somewhat different calibration strategy, so we 
will now describe in some detail the calibration procedures followed for each one.

\subsection{1982 observation}

The first observations were carried out on 1982, June 24 (epoch 1982.48) and they were first reported by Kwok \& Bignell (1984) --project code: BIGN. 
The correlator was setup to the continuum mode with a total bandwidth of 50 MHz. The observations consisted of one single scan on 
3C~48 (0134+329 --used as flux calibrator), 3C~119 (used as phase calibrator), and CRL~618, respectively.  The total observation time on the 
target source was 30~minutes. The flux scale was determined by assuming a flux density of 1.27~Jy for 3C~48. The 
bootstrapped flux density of the phase calibrator was 0.97~$\pm$~0.04~Jy. The flux density of CRL~618 was found to be 
205~$\pm$~11~mJy. Kwok \& Bignell (1984) did not explicitly quote an integrated flux, but mentioned that 
the sum of all clean components in their image was 214~$\pm$~7~mJy. Therefore our value is in good agreement with the previous 
measurement.  

\subsection{1983 observation}

The data of the second observation that we analyzed was obtained on 1983, October 9 (epoch 1983.77), and it has 
not been published before (project AK0094).  The observations were done in continuum mode at frequencies 5 and 22~GHz
with an effective bandwidth of 100 MHz. We performed calibration of the data at the two frequencies. The main flux 
calibrator was 3C~286 (1328+307) and the phase calibrator was J0403+260 for the observations at both frequencies. The total 
observation time on the target source at 22~GHz was $\sim$45~minutes, whereas the observation at 5~GHz lasted only one 
integration time of 10 seconds. The flux densities of the phase calibrator and the target source were scaled assuming that the 
flux density of 3C~286 is 7.5 and 2.57~Jy at 5 and 22~GHz, respectively. The bootstrapped flux of the source J0403+260 was
1.15~$\pm$~0.01~Jy, and 0.68~$\pm$~0.02~Jy at 5 and 22~GHz, respectively. For the observation at 22~GHz, after the initial 
calibration, we performed self-calibration of the data using a model-image from the previous epoch (1982.48).

\subsection{1990 observation}

The third observation was obtained as part of the project AM287 on March 29, 1990 (epoch 1990.24), and was published by Mart\'{\i}n-Pintado et al.\ (1993). 
The project was designed to investigate the distribution of ammonia in the surroundings of CRL~618, so the data were collected in spectral 
line mode. The frequency of observation was that of the NH$_3$ (2,2) inversion line at 23.72 GHz. In the original data, the quasar 3C~84 (0316+413) 
was used both as phase and flux calibrator (with an assumed flux at 22 GHz of 28 Jy). However, 3C~84 is now known to be variable. To 
estimate the flux of 3C~84 at the epoch of the CRL~618 observations, we searched the VLA archive for observing sessions in February, March, 
and April 1990, in which both 3C~84 and a standard flux calibrator were observed at $\lambda = 1.3$~cm. Four such sessions were found  
(Tab. 1), and were calibrated following standard procedures. The flux of 3C~84 obtained from these data appears to be 34~-~38 Jy (see Tab.\ 
1). An independent confirmation for such a high flux comes from Ter\"asranta et al.\ (1992) who monitored the 22 GHz flux of 3C~84 between 
January 1986 and June 1990. For the period February-April 1990, they reported fluxes of about 38 Jy, in excellent agreement with the values 
found here from the VLA data. The flux of 3C~84 appropriate for the date of the CRL~618 observations was estimated by interpolating the 
last two observations in Tab.\ 2 to be 37.7 Jy. Note that this is 35\% higher than the flux assumed by Mart\'{\i}n-Pintado et al.\ (1993). 
\begin{table}
\caption{Flux densities for 3C~84 in the year 1990}             
\label{table:1}      
\centering                          
\begin{tabular}{l c c c}        
\hline\hline                 
Date of observation~~~~~~& $\nu$ (GHz) & $S_\nu$ (Jy) & Project \\    
\hline                        
  15 February 1990 \dotfill & 22.2851 & 34.10 $\pm$ 0.55 & AR416 \\%
                 & 22.2320 & 33.64 $\pm$ 0.70 \\%
17 February 1990 \dotfill & 22.2851 & 34.38 $\pm$ 0.60 & AR216 \\%
                 & 22.2320 & 34.71 $\pm$ 0.86 \\%
04 March 1990  \dotfill  & 22.2851 & 36.02 $\pm$ 0.30 & AR216\\%
                 & 22.2305 & 36.38 $\pm$ 0.68 \\%
12 April 1990   \dotfill & 22.4851 & 38.42 $\pm$ 2.25 & TT001 \\%
                 & 22.4351 & 38.31 $\pm$ 2.24 \\%
\hline                                   
\end{tabular}
\end{table}
\begin{table*}[t!]
\caption{\label{t7}Parameters of the calibration of the observations at 22 GHz}
\centering
\begin{tabular}{llclccr@{}l}
\hline\hline
Date of observation~~~~~~~ &Flux calibrator & $S_\nu$\tablefootmark{1}& Phase calibrator& $S_\nu$\tablefootmark{2} &$rms$ Noise\tablefootmark{3}&&Beam\tablefootmark{4}\\%
\hline
(year) &(name)& (Jy)&(name)& (Jy)& (Jy~beam$^{-1}$)& &(~$^{\prime\prime}$~,~ $^{\circ}$~)\\%
\hline
24 June 1982 \dotfill & 3C48 \dotfill& 1.27&3C119\dotfill&0.97& 7.0$\times$10$^{-4}$& 0.11&$\times$0.09, $-$57\\%
9 October 1983 \dotfill & 3C286 \dotfill& 2.57&J0403+260\dotfill&0.68& 1.2$\times$10$^{-3}$& 0.10&$\times$0.08, $-$71\\%
29 March 1990 \dotfill & 3C84 \dotfill& 37.7 &3C84 \dotfill& 37.7& 6.5$\times$10$^{-4}$&0.13&$\times$0.09, $-$69\\%
22 December 1992  \dotfill  & 3C48 \dotfill&1.2& 3C84\dotfill & 26.6& 5.6$\times$10$^{-4}$&0.09&$\times$0.08, $-$42\\%
 & & & J0359+509 & 3.1& 5.6$\times$10$^{-4}$&0.09&$\times$0.08, $-$42\\%
3 August 1995   \dotfill & 3C286\dotfill &2.55 &3C84\dotfill& 19.4 & 1.0$\times$10$^{-3}$&0.11&$\times$0.10, $-$28\\%
2 May 1998   \dotfill & J0443+3441 \dotfill&0.35 &J0443+3441\dotfill& 0.35& 4.0$\times$10$^{-4}$& 0.14&$\times$0.10, $-$79\\%
9 July 2007   \dotfill & 3C48 \dotfill& 1.27&J04183+38015\dotfill& 5.2&1.3$\times$10$^{-4}$ & 0.12&$\times$0.09, $-$81\\%
\hline 
\end{tabular} \hspace{2cm}
\tablefoottext{1}{Assumed value for the primary flux calibrator.}
\tablefoottext{2}{Bootstrapped flux obtained using the AIPS task GETJY.}
\tablefoottext{3}{$rms$ noise in the image of the target source made with natural-robust weighting.}
\tablefoottext{4}{Size and position angle of the synthesized beam using natural-robust weighting.}
\end{table*}
\begin{table*}
\caption{Parameters of the calibration of the observation on epoch 1998.33}             
\label{table:3}      
\centering                          
\begin{tabular}{r@{}l c c c r@{}l}        
\hline\hline                 
&\hspace{-0.5cm}Frequency& Flux/Phase calibrator & $S_\nu$\tablefootmark{1} &$rms$\tablefootmark{3}&&Beam\tablefootmark{4}\\
 \hline 
&\hspace{-0.3cm}(GHz)& (name)& (Jy) &(Jy~beam$^{-1}$)&&(~$^{\prime\prime}$~,~ $^{\circ}$~) \\
 \hline 
~~~~~~~~~~1&.43  &  J0443+3441&0.70 &5.3$\times$10$^{-5}$ &2.0&$\times$1.6,    $-$66\\%
~~~~~~~~~~4&.89  &  J0443+3441&0.93 &3.5$\times$10$^{-5}$ &0.63&$\times$0.52,  $-$72 \\%
~~~~~~~~~~8&.44  &  J0443+3441&0.74 &4.7$\times$10$^{-5}$ &0.34&$\times$0.28,  $-$71 \\%
~~~~~~~~~14&.96 &   J0443+3441&0.50 &1.7$\times$10$^{-4}$ &0.20&$\times$0.16,  $-$77 \\%
~~~~~~~~~22&.49 &   J0443+3441&0.35\tablefootmark{2} &3.9$\times$10$^{-4}$ &0.14&$\times$0.10,  $-$79 \\%
~~~~~~~~~43&.31 &   J0443+3441&0.20 &3.1$\times$10$^{-4}$ &0.046&$\times$0.043, 30 \\%
  \hline                                   
\end{tabular}\hspace{6cm}
\tablefoottext{1}{Assumed value for the primary flux calibrator, obtained from the NRAO VLA calibrators webpage.}
\tablefoottext{2}{Interpolated value from the SED of the flux calibrator.}
\tablefoottext{3}{$rms$ noise in the image of the target source made with natural-robust weighting.}
\tablefoottext{4}{Size and position angle of the synthesized beam using natural-robust weighting.}
\end{table*}
The spectral line data were processed using standard AIPS procedures, including bandpass calibration. The central channels (at 
$-$80 km s$^{-1}$~$<$~v$_{\rm LSR} < -$25 km s$^{-1}$) show significant absorption by NH$_3$ (see also Mart\'{\i}n-Pintado \& Bachiller 
1992, Mart\'{\i}n-Pintado et al.\
1993, 1995). However, no such absorption is seen at v$_{\rm LSR} > -$20~km~s$^{-1}$. Thus, we averaged 17 absorption-free channels at 
v$_{\rm LSR} > -$20~km~s$^{-1}$ to produce a pseudo-continuum data set with an equivalent bandwidth of 6.6 MHz. This is significantly less than the 50 MHz 
bandwidth of the continuum observations obtained in 1982 (Sect.\ 2.1). Nevertheless, the noise level achieved is somewhat better in the 1990
than in the 1982 observations because the integration time was significantly longer, $\sim 2.5$~hours (Tab.\ 2). Several iterations of phase 
self-calibration were applied to the data to improve the quality of the phase calibration. The size of the emitting region is found to be 
\msec{0}{38}~$\times$~\msec{0}{13} comparable to the values reported by Mart\'{\i}n-Pintado et al.\ (1993) : \msec{0}{40}~$\times$~\msec{0}{12}. 
The total flux is found to be 293~$\pm$~15 mJy, significantly larger than the value of 189 $\pm$ 20 mJy reported by Mart\'{\i}n-Pintado et al.\ (1993). 
This, of course, largely reflects the difference in absolute flux calibration related to the updated flux density of 3C~84. 
\begin{table*}
\caption{Parameters of CRL~618 at $\sim$22~GHz}             
\label{table:3}      
\centering                          
\begin{tabular}{c c r@{}l c c c}        
\hline\hline                 
Epoch&Frequency&&$S_\nu$& Minor axis & Major Axis & P. A.\\    
 \hline 
year& GHz&&\hspace{-0.2cm}mJy& arcsec& arcsec & degrees\\
\hline 
1982.48  &  22.485&   205&$\pm$11          & 0.103$\pm$0.009& 0.333$\pm$0.015& 83$\pm$1\\%
1983.77  &  22.485&  210&$\pm$8& 0.115$\pm$0.012& 0.302$\pm$0.019             & 88$\pm$2\\%
1990.24  &  23.722&  300&$\pm$16          & 0.133$\pm$0.006& 0.383$\pm$0.013 & 81$\pm$1\\%
1992.98  & 23.872 &  368&$\pm$8& 0.146$\pm$0.004& 0.414$\pm$0.009             & 82$\pm$1\\%
1995.59  &  22.693&  401&$\pm$10          & 0.150$\pm$0.010& 0.458$\pm$0.022 & 82$\pm$1\\%
1998.33  & 22.485 & 415&$\pm$21          & 0.161$\pm$0.007& 0.455$\pm$0.014  & 82$\pm$1\\%
2007.52  &  22.485& 517&$\pm$10          & 0.185$\pm$0.001& 0.499$\pm$0.002  & 83$\pm$1\\%
     \hline                                   
\end{tabular}
\end{table*}
\begin{table*}
\caption{Flux of CRL~618 at $\sim$5~GHz}             
\label{table:4}      
\centering                          
\begin{tabular}{c cr@{}l c}        
\hline\hline                 
Epoch&Frequency &&$S_\nu$ & Telescope \\    
 \hline 
year& GHz &&\hspace{-0.2cm}mJy &  \\
\hline 
 1974.0 & 5.00& 6   &$\pm$3    &  \dotfill Cambridge 5-km\tablefootmark{1} \\%
 1978.2 &  4.80&14  &$\pm$2    & \dotfill  Effelsberg 100m\tablefootmark{2}\\%
 1980.2 & 4.89 &14.2&$\pm$0.5  & \dotfill  NRAO-VLA\tablefootmark{3} \\%
 1981.8 & 4.89 &16.2&$\pm$0.1  & \dotfill  NRAO-VLA\tablefootmark{4}\\%
 1982.5 & 4.89 &17  &$\pm$1    &  \dotfill NRAO-VLA\tablefootmark{5}\\%
 1983.8 & 4.89 &18.9&$\pm$1.8  &  \dotfill NRAO-VLA\tablefootmark{6}  \\%
 1986.5 & 4.89 &18.4&$\pm$1.8  & \dotfill  NRAO-VLA\tablefootmark{7}  \\%
 1991.0 & 4.89 &26  &$\pm$3    &  \dotfill NRAO-VLA\tablefootmark{8} \\%
 1998.3 & 4.89 &33  &$\pm$2    & \dotfill  NRAO-VLA\tablefootmark{6}\\%
\hline                                   
\end{tabular}\hspace{10cm}
\tablefoottext{1}{Wynn-Williams (1977)}
\tablefoottext{2}{Mross, Weinberger, \& Hartl (1981)}
\tablefoottext{3}{Kwok \& Feldman (1981)}
\tablefoottext{4}{Spergel, Giuliani, \& Knapp (1983)}
\tablefoottext{5}{Kwok \& Bignell (1984)}
\tablefoottext{6}{This work}
\tablefoottext{7}{Zijlstra, Pottasch \& Bignell (1989)}
\tablefoottext{8}{Knapp et al. (1995)}
\end{table*}
\subsection{1992 observation}

The fourth observation was obtained on 1992, December 22 (epoch 1992.98), and corresponds to the data published by Mart\'{\i}n-Pintado et
al.\ (1995) --project code: AM337. Since the primary goal of that observation was the study of the NH$_3$(3,3) transition, the observations were 
obtained in spectral mode at the frequency $\nu = 23.872$~GHz. To produce a continuum image, we averaged 20 line-free channels 
corresponding to v$_{\rm LSR} \geq$~$-$20~km~s$^{-1}$ (the spectra shown by Mart\'{\i}n-Pintado \& Bachiller 1992 clearly show that no absorption 
is present at those velocities). This resulted in a bandwidth of about 4 MHz. The observations consisted of scans on CRL~618, interleaved with 
scans on the calibrators J0359+509 and 3C~84 for phase-calibration purposes. The total observation time on the target source was 2.75~hours. 
One scan on the calibrator 3C~48  was done at the end of the observations.  Mart\'{\i}n-Pintado et al.\ (1995) 
used 3C~84 as amplitude calibrator assuming a flux at 1.3 cm of 28 Jy. The calibrator 3C~84 is known to have a structure that departs significantly 
from a point source at $\nu=22$~GHz. Besides, as mentioned above, this calibrator is also known to be variable. Because of these reasons,
we calibrated the amplitudes instead using the calibrator 3C~48, for which the flux density was assumed to be 1.2~Jy. We note that the bootstrapped
flux of 3C~84 obtained using 3C~48 as primary flux calibrator is 26.6~$\pm$~0.5~Jy, which is sightly smaller than the value assumed by
Mart\'{\i}n-Pintado (1995). The bootstrapped flux of 0355+508 was 3.1~Jy. After a few self-calibration iterations, we obtained an image of CRL~618 
from which we obtained its parameters. The integrated flux is 365 $\pm$~9 mJy, which is significantly larger than the value reported by 
Mart\'{\i}n-Pintado et al.\ (1995; 250~$\pm$~20 mJy; even though they assumed a larger value for the flux of 3C~84 than the bootstrapped value 
obtained from our calibration). This discrepancy could be mainly due to the difference of the amplitude-calibration procedures that we followed. 
However, our value is consistent with the monotonic increasing trend that the flux has followed during the last decades, which is also seen 
from the data at 5~GHz. This indicates that the value obtained by Mart\'{\i}n-Pintado et al. (1995) was affected in the calibration process.

\subsection{1995 observation}

The fifth observation was obtained in 1995, August 3 (epoch 1995.59) as part of the project AM0486. These data have not been published 
before. The observations were done in spectral line mode with a bandwidth of $\sim 6$~MHz centered at the frequency $\nu =22.693$~GHz. 
Since we did not find line emission across this bandwidth, we averaged the default  channel range (i.e. the center 75\%) to produce the 
continuum data set.  The primary flux and phase calibrators were  3C~286  and 3C~84, respectively. The assumed flux density for the primary 
flux calibrator was 2.55~Jy and 
the bootstrapped flux for the phase calibrator was  19.4~$\pm$~0.5~Jy. The total observation time on the target source was $\sim 1.2$~hours.
The data was initially calibrated using the quasars and subsequently the target source was self-calibrated in phase.

\subsection{1998 observation}

The data corresponding to the sixth observation was obtained in 1998, May 2 (epoch 1998.33) as part of the project AW048. These data have 
not been published before. The observations were done in continuum mode with a total bandwidth of 100~MHz. The array was split into two 
subarrays: VLA-A1 and VLA-A2, with 12 and 15 antennas, respectively. The subarray A1 included the antennas at the tips of the arms of the 
array, therefore it contained the longest baselines. Observations of CRL~618 at $\nu=8$ and 43~GHz were made using the subarray A1, whereas
the subarray A2 was used to carry out observations at $\nu=1.4$, 5, 15 and 22~GHz. We calibrated the data at the six different frequencies. 
The details of the calibration at the different frequencies are shown in Table 3. After the initial calibration, all the data corresponding to CRL~618 
was self-calibrated in phase.  \\
\begin{figure*}
   \centering
   \includegraphics[width=\hsize]{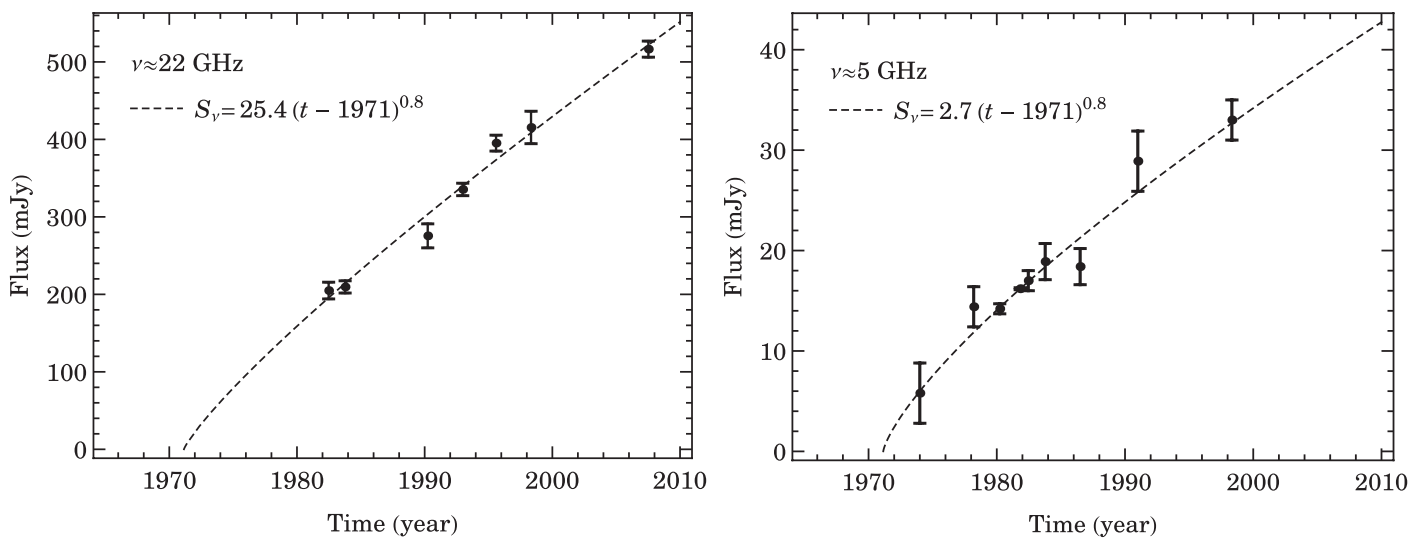}
      \caption{Increase of the flux density of CRL~618 at 22~GHz (left panel) and 5~GHz (right panel) as a function of time.  The dashed line is the 
      result of a simultaneous power-law fit to the data of the flux at 22~GHz and 5~GHz, and the size of the source at 22~GHz (Figure 2). 
      The numerical expressions of the fits are also shown in the panels. We set the starting time of the ionization as a common parameter of the fit (see main text).
              }
         \label{FigVibStab}
   \end{figure*}
\subsection{2007 observation}

The data from the last observation that we analyzed was obtained on 2007, July 9 (epoch 2007.52) in continuum mode with 100~MHz of equivalent 
bandwidth (project code: AL698). These observations were carried out using the fast-switching technique, which consists of rapidly alternating observations of the source 
and the phase calibrator (J0418+380) with a cycle time of 2 minutes. The total observation time on the target source was $\sim 35$~minutes. The 
primary flux calibrator was 3C~48 whose flux was assumed to be 1.27~Jy. The bootstrapped flux of the phase calibrator was  5.2~$\pm$~0.1~Jy. 
At the time of the observations, about half of the VLA antennas had already been upgraded to EVLA status.\\


\section{The evolution of the radio continuum of CRL~618}

Observations over the last decades have revealed that the emission of CRL~618 at mm and cm 
wavelengths exhibits changes over time (Kwok \& Feldman 1981; Kwok \& Bignell 1984; Mart\'\i n-Pintado et al. 1995; 
Knapp et al. 1995; S\'anchez-Contreras et al. 2004 and references therein). Particularly, at cm wavelengths 
the emission shows a more or less monotonic increase trend. Because the spectral index $\alpha$, defined 
as $S_{\nu}\propto\nu^{\alpha}$, of the radio continuum emission is very close to the value +2, the increase of 
the flux has been explained in terms of an expanding optically thick ionized region.  Knapp et al. (1995) presented 
a plot showing the increase of the radio continuum flux at $\sim 5$~GHz using data collected from 1974 until 1991; 
they extrapolated the value of the time that corresponds to flux equal zero and estimated that the ionization of the source began 
in about 1965.

From our analysis we have obtained measurements of the flux density of CR~618 at 
22~GHz at seven epochs ranging from 1982 until 2007 (see \S2). Additionally, we obtained the values 
of the flux at 5~GHz in two epochs to which we added the values of seven more 
epochs collected from the literature. The results are presented in 
Tables~4 and 5; in Figure~1 we have plotted the value of the flux at $\sim 22$ and $\sim 5$~GHz 
as a function of time, respectively. Since the frequency of the observation was not 
exactly the same for all the epochs (see Table~4), we scaled the values of the flux to a common frequency 
(22.485 and 4.885~GHz, respectively) for all the epochs, assuming that the spectral index has not 
changed considerably (see~\S4). From this figure we see that the flux at both frequencies has increased by a factor 
of more than two, since the first observations. The rise of the flux has been roughly linear with time, 
however, in order to also account for the growth of the size as a function of time, we have preformed a simultaneous 
power-law fit (see below) in which we have set as a common parameter the starting time of the ionization. 
\begin{figure}
   \centering
   \includegraphics[width=\hsize]{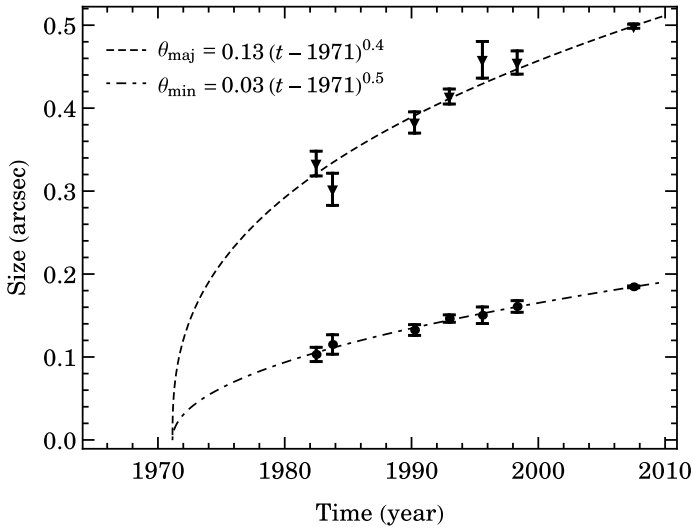}
      \caption{Increase of the size of CRL~618 at 22~GHz as a function of time. The dashed and dot-dashed lines are the fits to the 
      major and minor axes, respectively. The fit was obtained by performing a simultaneous power-law fit to the data of the flux at 
      22~GHz and 5~GHz (see Figure 1), and the size of the source at 22~GHz. The numerical expressions of the fits are also shown. 
      We set the starting time of the ionization as a common parameter of the fit (see main text).}
         \label{FigVibStab}
   \end{figure}

Even though images of this source at radio frequencies have been obtained and presented in previous works, so far 
the change of the size has been inferred only from the increase of the flux, and assuming that all the emission is optically thick 
(Kwok \& Feldman 1981; Spergel et al. 1983; Knapp et al. 1995). In this work we characterize the evolution of the size of 
ionized region by directly measuring the change of the brightness distribution from the images. Because the morphology of 
the source is not too complex at radio wavelengths, we have measured its size by fitting an elliptical gaussian model to its brightness distribution. 
From the fit we obtained the values of the major and minor axes, and the position angle of the major axis;  the results are presented in Table~4. 
In Figure~2 we plotted the increase with time of the size of the ionized region of CRL~618 at $\nu=22$~GHz, using the data from our 
elliptical gaussian fit to the images; the inverted triangles correspond to the major axis whereas the circles correspond to 
the minor axis of the source. In the same fashion as the flux, the size of the source has been experiencing a monotonic increase. 
We note that the major axis of the epoch 1983.77 is slightly smaller than that of the previous epoch, however, 
the level of noise of the phases in the visibilities of the epoch 1983.77 was significantly higher, resulting in a larger 
uncertainty in the determination of the size. A linear fit to the size as a function of time yields a zero-flux-time 
which corresponds to the beginning of the ionization, that differs by more than 30 years with respect to that obtained by 
performing the same type of fit to the data of the flux shown in Figure~1. This discrepancy is significant, even considering the errors of the 
fit. Because of this, we conclude that neither the flux nor the size have followed a linear increase with 
time. Thus, to a first approximation, we fitted a power-law of the form $a(t-t_{0})^{b}$ to all the data sets, setting as a common parameter 
the starting time of the ionization, $t_{0}$. The expressions of the fitted power-laws for the flux, major and minor axes are the following:
\begin{equation}
\begin{split}
\left[\frac{S_{\rm 22~GHz}}{\rm mJy}\right]=(25.4\pm10)\times\left(\left[\frac{t}{\rm year}\right]-(1971\pm2)\right)^{0.8\pm0.1}, \nonumber \\
\left[\frac{S_{\rm 5~GHz}}{\rm mJy}\right]=(2.7\pm1.3)\times\left(\left[\frac{t}{\rm year}\right]-(1971\pm2)\right)^{0.8\pm0.1}, \nonumber \\
\end{split}
\end{equation}
\begin{equation}
\begin{split}
\left[\frac{\theta_{\rm maj}}{\rm arcsec}\right]=(0.13\pm0.02)\times\left(\left[\frac{t}{\rm year}\right]-(1971\pm2)\right)^{0.4\pm0.1}, \nonumber \\
\left[\frac{\theta_{\rm min}}{\rm arcsec}\right]=(0.03\pm0.01)\times\left(\left[\frac{t}{\rm year}\right]-(1971\pm2)\right)^{0.5\pm0.1}. \nonumber \\
\end{split}
\end{equation}
\begin{figure*}
   \centering
   \includegraphics[width=\hsize]{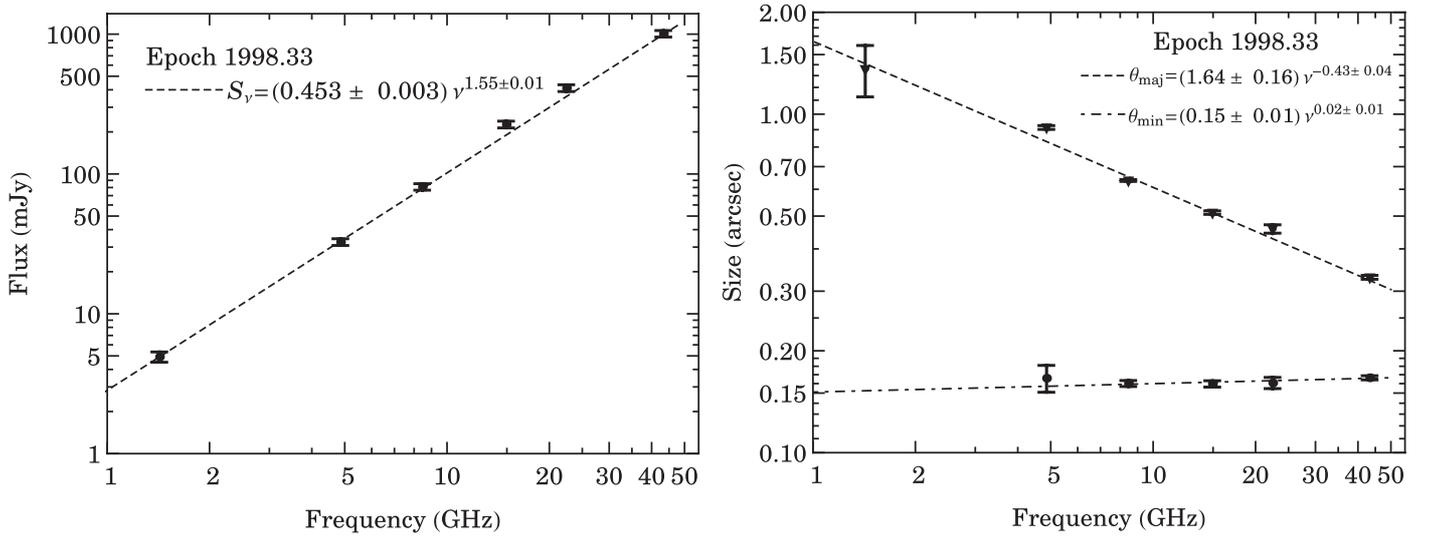}
      \caption{Dependence of radio flux and size of CRL~618 with Frequency. The data correspond to the observations carried out 
      in the epoch 1998.33. The minor axis of the source at $\nu=1.46$~GHz was not resolved, thus we only show the value of the 
      major axis. The dashed and dot-dashed lines represent least squares fits to the data, assuming a power law model. The numerical 
      expressions of the fits are also shown in the panels.}
         \label{FigVibStab}
   \end{figure*}
%
   %
From this results we notice that the increase of the flux is proportional to the increase of the area of the source, i.e. the exponent of the 
power-law fit of the flux is, within the errors, the sum of the exponents of the power-law fits of the minor and major axes. This result indicates
that indeed the rise of radio continuum is mainly due to the expansion of the projected area of the optically thick 
ionized region. The rate of expansion of the major axis is larger than that of the minor axis by a factor of nearly two, 
currently being 4.7~$\pm$~1.1 and 2.3~$\pm$~0.6~mas~yr$^{-1}$, respectively. If we consider a distance to the source $D=900$~pc and 
an inclination of the major axis with respect to the plane of the sky of $\sim 25^{\circ}$ (S\'anchez-Contreras et al. 2002), the measured 
expansion rates translate into deprojected linear velocities of 22~$\pm$~5 and 10~$\pm$~2 km~s$^{-1}$ for the major and 
minor axes, respectively. The fit gives a year for the beginning of the ionization about 1971, which is more recent 
that previously considered in other works. 
\begin{table*}
\caption{Parameters of CRL~618 for epoch 1998.33}             
\label{table:3}      
\centering                          
\begin{tabular}{c r@{}l c c c}        
\hline\hline                 
lines
Frequency&&$S_\nu$& Minor axis & Major Axis &P. A.\\    
 \hline 
GHz& &\hspace{-0.2cm}mJy& arcsec& arcsec& degrees\\
\hline 
1.46  &   5&$\pm$1&              \dotfill \tablefootmark{1}  &        1.360$\pm$0.236        & 89$\pm$3  \\%
4.89  &   33&$\pm$2 &                  0.166$\pm$0.015  &          0.914$\pm$0.012      & 87$\pm$1\\%
8.44  &   81&$\pm$4&                   0.160$\pm$0.003  &           0.637$\pm$0.004        & 82$\pm$1\\%
14.96  &   226&$\pm$13 &           0.160$\pm$0.003 &           0.512$\pm$0.006        & 82$\pm$1\\%
22.49  &   415&$\pm$21  &          0.161$\pm$0.006  &           0.458$\pm$0.013        & 82$\pm$1\\%
43.31  &   1007&$\pm$54 &         0.166$\pm$0.002  &           0.330$\pm$0.004        & 87$\pm$1\\%
     \hline                                   
\end{tabular}\hspace{10cm}
\tablefoottext{1}{The minor axis was not resolved at this frequency.}
\end{table*}
\section{The physical parameters of the ionized region}

In previous works the spectral index of the radio continuum  emission at cm wavelengths of CRL~618 has been 
considered to be $\alpha=+2$, or very close to this value (e. g. Kwok \& Bignell 1984). This implies that the emitting 
ionized region has a uniform electron density distribution. Given that the flux of this source evolves with time, one cannot
directly compare the measurements of the flux of one epoch with another. Thus, so far the only reliable value of the 
spectral index of the radio emission of this source is that derived from the measurements at three wavelengths obtained in 
quasi-simultaneous observations by Kwok \& Bignell (1984). Mart\'\i n-Pintado et al. (1988) performed a fit to the spectral 
energy distribution from cm wavelengths up to mm wavelengths. These authors concluded that the emission could not 
correspond to a HII region with uniform electron density. They suggested that the electron density should follow a power-law 
distribution of the form n$_{\rm e} = A\,r^{-2}$, which results in a spectral index $\alpha<+2$.  Indeed, a power-law fit to 
the data of Kwok \& Bignell (1984) gives a spectral index  $\alpha=+1.71\pm0.01$. 

Among the data that we have analyzed in this work, there are simultaneous observations of the radio continuum of CRL~618 at six 
different frequencies (see \S 2). We measured the flux density of the source at each frequency. Additionally, for each frequency 
we fitted an elliptical gaussian model to the brightness distribution to measure the major and minor axes, as well as the position angle 
of the major axis of the emitting region; the results are presented in Table~6. In Figure~3 we show the integrated flux and the 
major and minor axes of CRL~618 as a function of the frequency. A power-law fit to the data of the flux yields a spectral index in 
the frequency range from 1 to 43~GHz of $\alpha =+1.55 \pm  0.01$.  This value is smaller than the value derived from the 
data obtained by Kwok \& Bignell (1984), which implies that the spectral index has been decreasing over time. However, this 
spectral index results from averaging the emission over the whole source. In order to obtain more information about the distribution 
of the ionized gas (i.e. the electron density) we combined the emission at five different frequencies, 5, 8.4, 14, 22 and 43~GHz, 
to make a map of the spectral index across the source; the result is shown in Figure 4. In this figure we see that the spectral 
index of the emission ranges from values near zero, to values up to +2 for different parts of the source. The highest values 
correspond to a band that crosses the source approximately in the North-South direction, and it exhibits a steep gradient in the 
East-West direction.
\begin{figure*}
   \centering
   \includegraphics[scale=0.65]{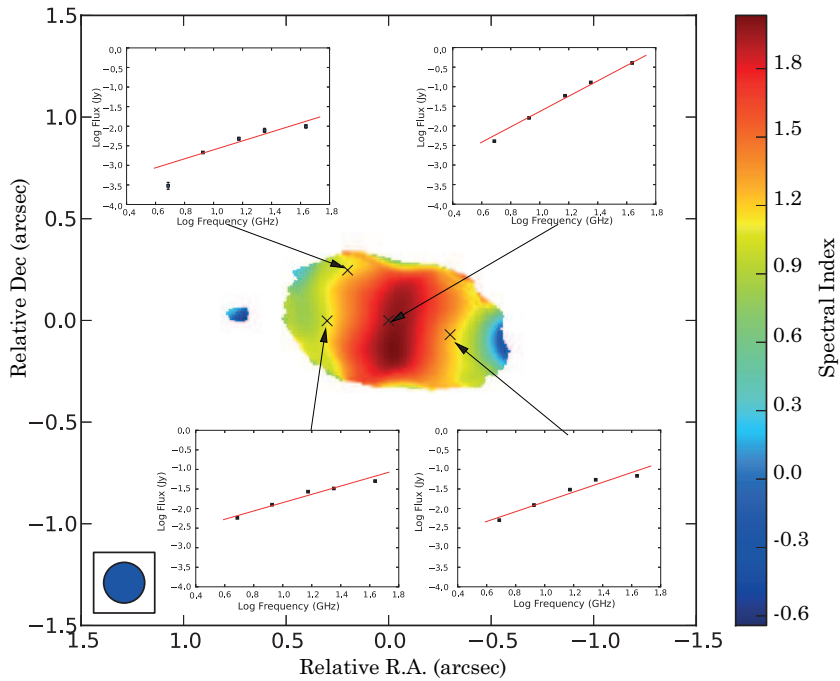}
      \caption{Map of spectral index of CRL~618 made using the emission at five frequencies from 5 to 43~GHz. 
      The inner panels show the fit to the SED for particular points of the source. A common synthesized beam of 
      0$\rlap{.}^{\prime\prime}$2$\times$ 0$\rlap{.}^{\prime\prime}$2 was used to make the map. }
         \label{FigVibStab}
   \end{figure*}

For the case in which the depth and electron temperature of the ionized region are constant across the source (below we show 
that this assumption is a good approximation), this map basically traces the electron density of the plasma. Thus, this spectral 
map shows that the electron density reaches a maximum on a band in the central region and decreases outwards in the 
East-West direction, which is the direction of the bipolar lobes seen at optical wavelengths. This gradient of the electron density 
is also revealed by the dependance of the major axis with frequency, which is plot in the right panel of Figure 3. From this plot we 
can see that while the minor axis remains basically constant for all the frequencies, the major axis shows a clear dependance 
on the frequency, $\theta_{\rm maj} \propto \nu^{-0.43\pm0.04}$. This behavior of the major axis can be explained in 
terms of the emission becoming optically thin at different distances from the star. Once the emission becomes optically thin, the brightness of the 
emission decreases rapid as a function of the optical depth. Reynolds (1986) demonstrated that for a partially optically thick 
ionized outflow whose width, depth and electron temperature are constant, the total integrated flux is proportional to the 
product of the size of the optically thick region and the frequency squared: $S_{\nu} \propto \theta_{\rm thick} \cdot \nu^{2}$.  
Since we have found from our observations (right panel of Figure 3) that  $\theta_{\rm maj}  \propto \nu^{-0.43}$,  assuming that 
the size of the major axis is basically the size of the optically thick region, we have that $S_{\nu} \propto \nu^{2-0.43}=\nu^{1.57}$, 
which is exactly the behavior of the total integrated flux that we see in the plot on the left panel of Figure~3. Therefore, we can 
consider that the width, depth and electron temperature of the ionized region are basically constant and that $\theta_{\rm maj}$ 
corresponds to the size at which the emission becomes optically thin, $\tau_{\nu} \approx 1$.\\

The optical depth along one line-of-sight path of a region of ionized gas can be approximated by the following expression:
\begin{equation}
\tau_{\nu}=8.235\times10^{-2}\,\left[\frac{EM}{{\rm cm}^{-6}\, {\rm pc}}\right]\left[\frac{T_{\rm e}}{\rm K}\right]^{-1.35}\left[\frac{\nu}{\rm GHz}\right]^{-2.1}, 
\end{equation} 
where $T_{\rm e}$ is the electron temperature and $EM$ is the emission measure, defined as $EM=\int n_{\rm e}^{2}{\rm d}L$, where 
$n_{\rm e}$ is the electron density and d$L$ is the differential path length (Altenhoff et al. 1960). If we assume that the electron density remains more or less 
constant along one line-of-sight path, we can write the emission measure as $EM=n_{\rm e}^{2}\,L$, where $L$ is the total depth of the 
source in the line of sight path. Thus, for a given frequency, from the expression above we can calculate the minimum electron density 
necessary for the gas to become optically thick along one line-of-sight path as follows.\\ 

Assume that the density can be expressed as a power-law of the distance to the star of the form $n_{\rm e} = A\,r^{\,\, \rm q_{n}}$.  
Consider the relation between the major axis and the frequency that we obtained from the fit shown in Figure 6, $r_{\tau=1} = 0.82\,\nu^{-0.4}$, 
where we have used $ r_{\tau=1} = \theta_{\rm maj} /2$.  Solving for  $n_{\rm e}$ in (1) and assuming that $L$ and $T_{\rm e}$ do not depend 
on the distance from the central star, nor on the frequency of observation, we can obtain the following relation:
\begin{equation*}
 A\cdot(0.82\, \nu^{-0.4})^{\rm q_{n}}=3.48\,(T_{\rm e}^{1.35}\,\nu^{2.1}\,L^{-1})^{1/2}.
\end{equation*} 
By equalizing the exponent of the frequency on both sides of the equation, we find that q$_{\rm n} =-2.4 \pm 0.2$. 
Thus, the expression for the coefficient $A$ is
\begin{equation*}
A=2.3\,(T_{\rm e}^{1.35}\,L^{-1})^{1/2}.
\end{equation*} 
We can further assume that the depth of the region is equal to the width of the minor axis, i.e. $L=\theta_{\rm min}$, 
and that the electron temperature is $T_{\rm e}$=13000~K (Mart\'\i n-Pintado et al. 1988). Using these values 
to calculate $A$ and using the value of  q$_{\rm n}$ obtained above, we can write the expression of the electron density 
along the direction of the major axis as:
\begin{equation}
 \left[\frac{n_{\rm e, \tau=1}}{\rm cm^{-3}}\right] = (4.9\pm0.9)\times10^{4}\left[\frac{r^{-2.4\pm0.2}}{\rm arcsec}\right], 
\end{equation} 
where we have assumed a distance to the source $D =900$~pc. In Figure 5 we show a plot of the electron density along the major axis as a function 
of radius. The grey region corresponds to the errors in the 
estimation of the parameters $A$ and q$_{\rm n}$ .  This plot shows that the electron density ranges from 
n$_{\rm e} \sim10^{5}$~cm$^{-3}$ to more than  n$_{\rm e} \sim 10^{6}$~cm$^{-3}$ across the source, 
with the highest value corresponding to the electron density in the equatorial band seen in Figure~4. As a matter of fact, 
the value of the electron density in this region represents just a lower limit, since the emission is optically thick.  
The solid line in Figure 5 indicate the range of size and density within which we have observational data (delimited by the 
dot-dashed lines). The dashed lines are extrapolations of the electron density for other values of the radius. Following 
the same analysis but using the data obtained by Kwok \& Bignell (1984), we estimated the corresponding 
values of the parameters of the power law of the electron density for that observation epoch. We found a very similar 
exponent of the power law, q$_{\rm n}=-2.4$, but a smaller coefficient $A=(4.2\pm0.1)\times10^{4}$, suggesting a 
smaller value for the electron density in the past (see below).\\
\begin{figure}
   \centering
   \includegraphics[width=\hsize]{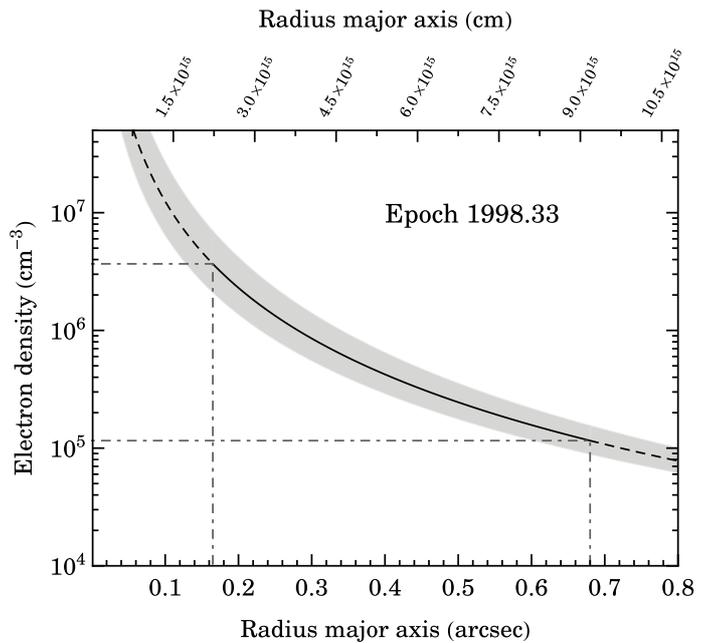}
      \caption{Dependence of the electron density with the radius for the compact HII region of CRL~618. The solid line 
      indicates the region where we have measurements of the size of the major axis as a function of frequency (see Figure 4). 
      The dashed lines are extrapolations of the fit following the same power law dependance. The gray region represents the 
      uncertainty region due the error of the fitted parameters. The horizontal/vertical dot-dashed lines indicate the extreme values 
      of the electron density and size of the source, corresponding to the observing frequencies: $\nu=43.31$ and 1.46~GHz (see Table~6).}
         \label{FigVibStab}
   \end{figure}
\begin{figure}
   \centering
   \includegraphics[width=\hsize]{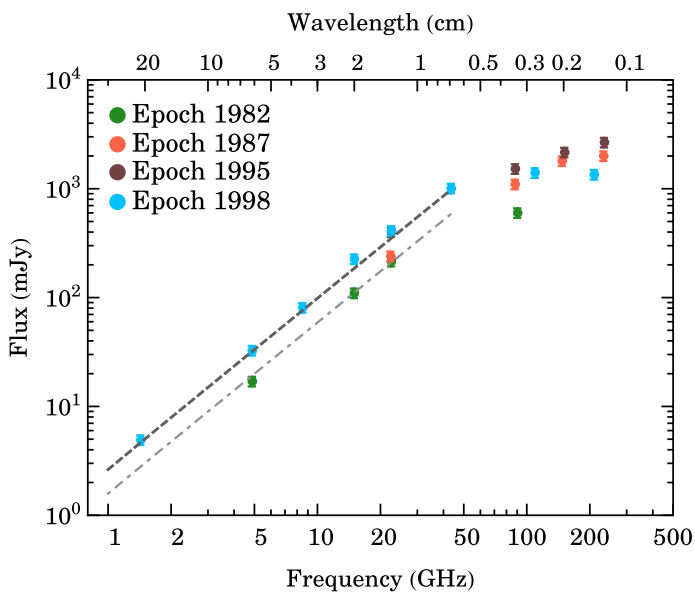}
      \caption{Spectral energy distribution of the radio continuum emission of CRL~618. The colors 
      indicate the data from the same epoch, as it is shown in the upper left corner of the Figure. The 
      dashed line corresponds to calculations of a model for the epoch 1998 (see main text). The dot-dashed 
      line corresponds to calculations of a model for the epoch 1982 (see main text). The data at  wavelengths $<$5 mm were 
      taken from Kwok \& Feldman (1981; extrapolation to epoch 1982); Mart\'in-Pintado et al. (1988; epoch 1987), Reuter et al. (1997; epoch 1995); 
      S\'anchez-Contreras \& Sahai (2004; extrapolation to epoch 1998) }
         \label{FigVibStab}
   \end{figure}

On the other hand, the size of the minor axis does not exhibit dependance on the frequency. This indicates that the emission is 
optically thick at all observed frequencies in this direction, which can be directly seen in Figure 4. From Figure 5 we see that the 
electron density in this region must be higher than $\sim 3 \times 10^{6}$~cm$^{-3}$. Observations of the molecular emission 
toward CRL~618 have revealed the presence of a dense torus-like structure that extends perpendicular to the bipolar lobes 
(S\'anchez-Contreras \& Sahai 2004). The density of the inner radius of the torus has been estimated to be of the order of 
10$^{6}$~cm$^{-3}$, which is in perfect agreement with our results. Since the HII region is ionization-bound in the direction 
of the minor axis, the increase of its size in this direction is mainly due to the propagation of the ionization front. 
Our observations can be successfully modeled by analytical calculations of the rate of expansion for an HII region, however, 
we will present those results somewhere else.
\\

In order to corroborate that the electron density has been increasing with time in the past, we performed calculations of the spectral energy 
distribution at two different epochs using a model similar to that of Mart\'\i n-Pintado et al. (1988). In our calculations, we used the electron density 
profile given in eq.2 (epoch 1998.33), as well as the profile obtained by using the parameters derived from the observations 
of Kwok \& Bignell (1984; epoch 1982.48). 
The geometry of the emitting region was assumed to be cylindrical, with the diameter of the cylinder equal to the measured minor axis. 
We assumed an electron temperature $T_{\rm e} =13000$~K and a distance to the source $D =900$~pc. The results of the calculations are 
shown in Figure 6 as a dashed and dot-dashed lines. The data from the different epochs are coded with colors. The calculations fit the data 
very well. This result confirms that the assumptions to estimate the electron density profiles are reasonable, which supports the idea that 
the electron density has been increasing with time in the last decades. 

\section{Discussion}

\subsection{An ionized wind in CRL~618?}

Evidence of the increase of the radio continuum emission in CRL~618 was reported for the first time by Kwok and Feldman (1981).
These authors showed that the flux doubled its value between 1977 and 1980. They explained such rapid rise of the radio flux as 
a result of the expansion of the ionization front into the neutral circumstellar envelope. In subsequent works it was reported that 
the increase of the radio flux seemed to have declined considerably, or even halted (Kwok and Bignell 1984; Mart\'\i n-Pintado et al. 1988; 
Sahai, Claussen, \& Masson 1989; Mart\'\i n-Pintado et al. 1993; Mart\'\i n-Pintado et al. 1995). This was attributed to a deceleration 
of the ionization front (e.g. Spergel, Giulliani \& Knapp 1983). 

The flux of the emission of CRL~618 at mm wavelengths has also shown an increasing trend since 1977.  A compilation of the flux at 
mm wavelengths as a function of time since 1977 until 2002 is presented in Table 2 of S\'anchez-Contreras et al. (2004).  In 
the plot of Figure~6 we included the flux at mm wavelengths of four epochs. The flux exhibits an increasing trend for the first three epochs but it decreases 
in the last epoch (1998). Mart\'\i n-Pintado et al. (1988) found that this source had increased its flux at 3.4 mm by a factor of 3.5 since 1977. 
These authors pointed out that for a power-law distribution of the electron density, n$_{\rm e} \propto r^{-2}$, the increase of the flux at mm wavelengths cannot be 
explained in terms of the expansion of the ionized region. Thus, they concluded that, to explain the increase of the flux density, one has 
to assume density inhomogeneities produced by changes in the mass-loss rate. Our observations also suggest that the density increased 
in the past. Now we will show that the increase of the electron density can also explain the growth of major axis of the HII region of CRL~618. 
\begin{figure*}
   \centering
   \includegraphics[scale=0.55, angle=-90]{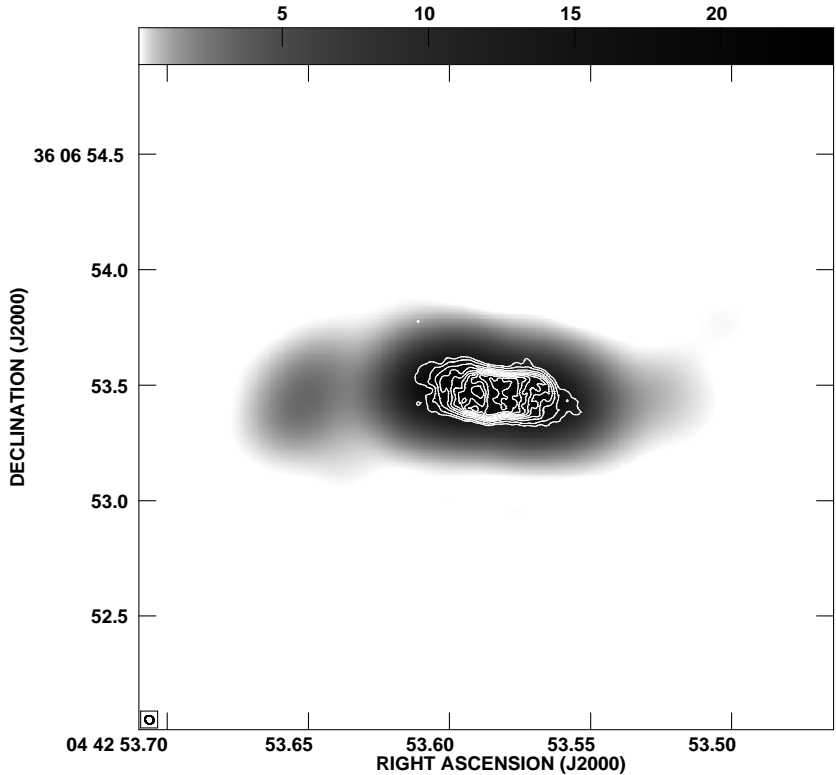}
      \caption{ Radio continuum emission of CRL~618 at 43~GHz (contours) superimposed on a gray scale image of the radio continuum emission at 8.4~GHz. 
     The values of the contours are ($-$5, 5, 10, 20, 30, 40, 50, 60, 70, 80, 90, 100)~$\times$~2.4$\times$10$^{-4}$~Jy~beam$^{-1}$, the $rms$ noise of the image 
     at 43~GHz. The size of the beam at 43~GHz is  $\sim 0 \rlap{.}^{\prime\prime}$043$\times$0$\rlap{.}^{\prime\prime}$046 with P.~A.$=30^{\circ}$, and it is 
     shown in the bottom left corner of the image. The gray scale uses the logarithmic intensity transfer function $I^{\prime}={\rm Log}_{10} (1+9\cdot I)$, where 
     the intensity, $I$, covers the range of values from 0 to 2.38$\times$10$^{-2}$~Jy~beam$^{-1}$. The size of the beam at 8.4~GHz, not shown in the image, is
     $\sim 0 \rlap{.}^{\prime\prime}$34$\times$0$\rlap{.}^{\prime\prime}$28 with P.~A. $=-71^{\circ}$.}
         \label{FigVibStab} 
   \end{figure*}

We have shown that as of 2007 the flux and the size of the source at 22~GHz have been continuously increasing. The radio images 
of CRL~618 at 5~GHz taken on early 1980, as well as those obtained in this work (epoch 1998), show radio continuum emission extending 
$\gtrsim 0 \rlap{.}^{\prime\prime}7$. This indicates that by the year 1980 the ionization front had expanded at least up to that distance. Therefore, 
the advance of the ionization front cannot explain the observed expansion of the major axis of the ionized region at 22~GHz, which extends only 
$\lesssim 0 \rlap{.}^{\prime\prime}5$ (see Figure~2). In order for the optically thick ionized region to expand, as our results indicate, it is 
necessary that its optical depth at larger radii increases with time. From (1) we see that the optical depth at a fiducial frequency on a line-of-sight 
path depends on the emission measure and the electron temperature. In \S4 we found that assuming a constant electron temperature and 
depth of the source was a good approximation to explain the spectral index and the dependance of the major axis as a function of frequency. 
Furthermore, from our observations we found that the peak brightness of the emission does not vary in more than 10\% from epoch 1982.48 to 
Epoch 2007.52. For an optically thick region, the peak brightness depends only on the electron temperature. Therefore, we do not expect 
the electron temperature to have changed more than 10\% throughout the observing epochs. Thus, the contribution to the optical depth due 
to a variation of the electron temperature should be of the order of 20\%. From Figure~2 and Table~4 we see that minor axis increased by a 
factor of $\sim 1.8$ since the first epoch. If we assume that the depth of the source is equal to the minor axis, and since the emission measure 
is proportional to the depth of the source, we have that for a given line-of-sight path the emission measure should have increased by a 
factor of $\sim 1.8$ throughout the observing epochs. If we consider the line-of-sight paths that correspond to the edges of the source 
($r_{2007.52} =0 \rlap{.}^{\prime\prime}5$ and $r_{1982.48} = 0 \rlap{.}^{\prime\prime}3$), for the epochs 1982.48 and 2007.52, 
using our result that the electron density is proportional to $r^{-2.4}$, we have that the ratio of electron densities should be 
(0.5/0.3)$^{-2.4} \approx 0.3$. Thus, while the increase of the depth of the source accounts for a 
factor of 1.8 in the increase of the optical depth, the decrease of the density with distance introduces a factor of $0.3^{2}=0.09$, which 
results in an effective growth of $\Delta \tau_{\nu} \sim 20$\%. From this we conclude that the increase of the minor axis and temperature 
alone cannot account for a significant increase of the optical depth and the main factor responsible for the increase the optical depth at a given 
distance from the star is the rise of the electron density. 

From a detailed analysis of the central region of CRL~618 at optical wavelengths, S\'anchez-Contreras et al. (2004) concluded that the gas 
in this region must be completely ionized. Therefore, the ionization fraction is practically unity 
and an increase of the ionizing photons does not result in an increase of the free electrons. 
The only way of increasing the electron density is by feeding in ionized gas with higher density. Thus, our results support the idea that 
CRL~618 is still expelling gas in the form of an ionized wind and whose mass-loss rate has been increasing with time.
We used the equation of mass continuity in order to reproduce the electron density profile shown in Figure~5. The best fit to our 
data gives an expansion velocity of the ionized gas of $\sim 22$~km~s$^{-1}$ and an increase of the mass-loss rate from 
$\sim 4 \times 10^{-6}$ to $\sim 6 \times 10^{-6}$~M$_{\odot}$~yr$^{-1}$ in the $\sim 130$~years previous to 1998. This value of the 
mass-loss rate agrees with that obtained by Mart\'in-Pintado et al. (1988) within a factor of two, once the distance to the source is scaled 
to $D=900$~pc.\\

\subsection{The origin and morphology of the ionized wind}

The beginning of the ionization of the circumstellar envelope at this very early stage of the evolution of the post-AGB phase 
suggests that CRL~618 descends from a relatively high mass star. The central star of CRL~618 has been classified to be 
spectral type B0 with a stellar temperature 
$T_{\star} \gtrsim 32000$~K (Westbrook et al. 1975; Kaler 1978; Schmidt \& Cohen 1981). S\'anchez-Contreras et al. (2002) found 
evidence of CIII / CIV lines corresponding to the W-R bump, which is observed in the majority of carbon dominated Wolf-Rayet-type, 
[WC], central stars of PNe. 
The presence of a [WC] central star might explain the high 
mass-loss rate observe in this source. It also might explain why the mass-loss rate was increasing during the last century but 
it decreased recently. The [WC] stars of post-AGB objects are known to experience periods of intense mass-loss. 
It could be possible that the central star of CRL~618 underwent a period of high mass-loss rate followed by a decrease 
of the mass ejection. Sometime around 1971 the electron density at the very base of the wind became low enough so that 
the ionizing photons could escape, ionizing the entire wind. The sudden increase of the flux seen in the years before 1980 
could have been due to the advance of the ionization front. Once the wind was completely ionized, the increase of the size 
of the HII region has been mainly due to the expansion of the higher electron density parts of the ionized wind at 
constant velocity. Therefore, we consider that we are witnessing the birth of a planetary nebula in CRL~618. 

In Figure~7 we present a radio image of CRL~618 at 43~GHz (shown in contours) superimposed on an gray-scaled image 
at 8.4~GHz. The size of the synthesized beam of the 43~GHz image is $\sim 0 \rlap{.}^{\prime\prime}$045, which makes it the radio image 
of CRL~618 with the highest angular resolution available so far. This image shows that the HII region has very well defined boundaries, parallel 
to each other, in the direction of the minor axis. The images at 5,  8.4, 15 and 22~GHz reveal that the source exhibits a subtle 
S-shape delineated by two faint extensions (see gray-scale image in Figure 7). Particularly, the extension toward the left  appears 
to be disconnected from the main part of the HII region. This structure had not been detected previously, most likely due to the lack of 
sensitivity. Since it lies along the direction of the major axis, it might be an enhancement of density of the ionized stellar wind, 
suggesting that there has been an episode of discrete mass ejection. The S-shape morphology of the ionized region hints at
possibility of the presence of some kind of precessing mechanism, which would lead to the suggestion of a binary system. 
In fact, if one of the members of the binary system is a white dwarf and the other is a mass losing AGB star, it could be possible 
to explain our observations in terms of the white dwarf ionizing the wind from the AGB star whose mass-loss rate had been 
increasing in the last century and dropped just recently. 

Given the elongated morphology in the East-West direction of the wind, one might wonder whether this is because the wind is intrinsically 
collimated in this direction (i.e. a bipolar ejection), or it is spherical but suffers some kind of confinement along this direction. 
The optical images obtained with the Hubble Space Telescope reveal the presence of bullet-like structures located at the tips of 
the finger-like lobes. Even though they are at different distances from the central star, Balick et al.~(2013)  have 
found that they all have roughly the same kinematical age, $\sim 100$~years. Thus, it seems that around a century ago there was an event 
in which material was expelled with large velocities, creating cavities within the steady expanding wind of the AGB phase. This event 
was followed by the onset of a stellar wind whose mass-loss was increasing with time until recently. In principle, the stellar wind does 
not need to be collimated intrinsically, but it expands preferentially in the direction of the cavities, due to the relative lower density in 
this direction. Thus, the full photoionization of the wind in the East-West direction seems to be directly related to the event that occurred 
$\sim 100$~years ago, which created the cavities. Had not these cavities been created, the ionization of the envelope would be constrained 
to a radius equal to the size of the minor axis in all directions. It seems that in the case of CRL~618 the configuration to 
form a bipolar planetary nebula was created only a century before the ionization of the CSE began. Therefore, the post-AGB phase of 
this source was extremely short.

\section{Conclusions}

   \begin{enumerate}
      \item We have traced the increase of the radio continuum flux at 5 and 22~GHz for a period of $\sim$26 years. The flux exhibits a 
      monotonic growth with time up to 2007. 
      \item We have measured the growth of the size directly from the brightness distribution of the emission. The major 
      axis is expanding twice as fast as the minor axis. 
       The increase of the projected area of the ionized region corresponds to the increase of the radio flux, indicating that the emitting 
       region is basically optically thick. 
       \item The ionization of the material around CRL~618 began some time around 1971. In the first years the growth of the ionized region 
       was steeper, and recently it became roughly linear. 
       \item While the major axis of the ionized region exhibits a dependance with frequency, the size of the minor axis is the same at all 
       observing frequencies. This behavior is explained in terms of the presence of an electron density gradient in the direction of the major axis. 
       The HII region is ionization-bound in the direction of the minor axis, indicating a much higher density of the material in this direction. 
       \item The density of the material in the direction of the major axis has increased in the last century, although in the recent decades 
         it seems to have experienced a drop. The decrease of the density close to the central star explains the beginning of the ionization 
         around 1971 and the observed expansion in the direction of the major axis. This event marked the beginning of the planetary 
         nebula phase for CRL~618, implying a very short time for its post-AGB phase, $\sim 100$~years.    
       
       \end{enumerate}

\begin{acknowledgements}
This research was supported by the Deutsche Forschungsgemeinschaft (DFG; through the Emmy Noether Research grant VL 61/3-1). 
L.L.\ acknowledges the financial support of DGAPA, UNAM and CONACyT, M\'exico. J. P. F acknowledges the Spanish Ministerio de 
Educaci\'on y Ciencia for funding support through grants ESP 2004-665 and AYA 2003-2785 and the ``Comunidad de Madrid'' government 
under PRICIT project S-0505/ESP-0237 (ASTROCAM). This study has been supported in part by the UNAM through a postdoctoral fellowship 
and the European Community's Human Potential Program under contract MCRTN-CT-2004-51230, ``Molecular Universe''. The authors 
are grateful to Bruce Balick for valuable discussions and suggestions.    
\end{acknowledgements}

\end{document}